# Thermal and non-thermal radiation of rotating polarizable particle moving in an equilibrium background of electromagnetic radiation

A.A. Kyasov and G.V. Dedkov[1]

Nanoscale Physics Group, Kabardino-Balkarian State University, Nalchik, Russia

**Abstract –**A theory of thermal and nonthermal radiation in a vacuum background of arbitrary temperature generated by relativistic polarizable particle with spin is proposed. When the particle rotates, radiation is produced by vacuum fluctuations even in the case of zero temperature of the system. In the ultrarelativistic case, the spectral-angular intensity of radiation is concentrated along the velocity of particle and vacuum. At finite temperatures of particle and vacuum, the particle temperature (in its rest frame) rather quickly acquires an equilibrium magnitude depending on the velocities of rotation and uniform motion and the background temperature. This equilibrium temperature determines the intensity of radiation. The dynamical slowing down takes a very long time until the kinetic energy of uniform motion and rotation is converted into radiation.

## 1. Introduction

Nowadays, stationary but dynamically and/or thermally non-equilibrium physical situations including various important physical systems like a small particle (an atom) in a vacuum background, two particles or the bodies in relative motion or out of thermal equilibrium, etc. have attracted continuously growing interest of many researchers [1—21]. Dynamically non-equilibrium situations involve the systems in relative motion [3—8,12,16,17,21] and rotating ones [9,14,15,18—20]. Close relation between the aforementioned non-equilibrium problems allows one to consider different parts of these systems as being in local equilibrium and to use a generalized form of the fluctuation-dissipation relations together with general principles of relativistic invariance [2,3,9,12,14--17]. On the other hand, similarity of the non-equilibrium problem statement for the bodies in relative motion or having different temperatures stems from the point of view of relativistic thermodynamics [22].

However, unlike the long-standing issues involving the forces of fluctuation-electromagnetic attraction/friction and heat exchange, much less attention has been paid to the radiation produced by rotating/moving neutral polarizable bodies [9,18--21]. At total dynamical and thermal equilibrium, a neutral particle embedded in vacuum emits and absorbs an equal number of thermal photons. This equilibrium is violated for moving/rotating particles, when the spectra of radiation (absorption) are dependent on translational and rotational velocities of motion. The spectral-angular distribution of long-wavelength radiation of such particles [21] differs significantly from the radiation of a moving (rotating) macroscopic body (black body, for example) [23, 24], since the particle radius is much lower than the Wien wavelength of thermal radiation.

---

[1] Corresponding author e-mail: gv_dedkov@mail.ru

For a spinless relativistic particle in a heated background (thermalized photonic gas) the corresponding radiation spectrum was calculated in our previous work [21]. The aim of this paper is to study the effect of combined action of uniform motion and rotation on the particle emission spectrum. We assume that the particle and vacuum are characterized by different local temperatures well defined in their respective frames of reference. If the rotation velocity of the particle is nonrelativistic and its radius is small compared to the characteristic Wien wavelength of thermal photons, then it can be treated as a point-like rotating fluctuating dipole. We obtain a full set of equations describing the dynamics and properties of radiation emitted by a moving spin-possessing particle. In particular, we formulate the condition of particle radiation at zero temperature of the entire system. This condition meets the condition of Zel'dovich superradiance [25, 26] where a rotating cylinder made of absorbing material is capable of amplifying certain modes of impinging electromagnetic radiation, i.e. an additional emission of the particle appears. In the case of nonmagnetic rotating particle with zero linear velocity the expression for the integral intensity of radiation was recently obtained in [20] and coincides with our result. A striking new finding is that the intensity is the same in the case of relativistic translational motion, as well as upon parallel and perpendicular mutual orientation of the vector of linear velocity and angular rotation axis. This is a clear manifestation of the fact that radiation of rotating particle appears as a result of its interaction with vacuum fluctuations of electromagnetic field. We have calculated the corresponding angular and frequency spectra in the particular case of Drude-like polarizability, and the general characteristics of radiation in the case out of thermal equilibrium.

In addition to their fundamental significance, the obtained results may be of interest upon creating new sources of directional microwave radiation, particle trapping in cavities, evolution of gas-dust clouds in cosmic space, etc.

The paper is organized as follows. In Sec. 2 we elaborate the arguments for the long-wavelength thermal and nonthermal radiation by relativistic polarizable particle with spin and arbitrary dielectric properties, which moves inertially in a vacuum background of a certain temperature. In Sec. 3 we obtain the basic formulas for the intensity of emission and absorption of the long-wavelength radiation depending on the linear and angular velocity of the particle, its temperature and the temperature of vacuum. In Sec. 4 we discuss the general characteristics of nonthermal radiation which is emitted at zero temperature of the particle and vacuum background. Section 5 is devoted to the calculation of nonthermal and thermal radiation at various conditions. Numerical calculations are performed in the case of particle polarizability for good conductors in the low-frequency limit of Drude model. Section 7 summarizes the conclusions. Appendix A contains an identity of the Joule dissipation integral in the co-moving reference frame, the particle heating rate in its own frame of reference and the work spent onto the braking of particle rotation in the co-moving frame. This allows one to obtain the equation for the rate of the change of particle temperature in its rest frame. Appendix B contains the basic results in configuration with perpendicular mutual orientation of spin and velocity vector.

## 2. Problem statement and general relations

Consider a small particle of radius $R$ and temperature $T_1$ uniformly moving with velocity $V$ through an equilibrium background radiation with temperature $T_2$ (Fig. 1, reference frame $\Sigma$) and rotating with angular velocity $\vec{\Omega} = (\Omega, 0, 0)$ in a co-moving reference frame $\Sigma'$. The third frame of reference $\Sigma''$ denotes the rest frame of the particle rigidly rotating relative to the system $\Sigma'$ with velocity $\vec{\Omega}$. Since the latter velocity is assumed to be nonrelativistic, this means that $\Omega R / c << 1$. We also assume that the conditions $R << \min(2\pi \hbar c / k_B T_1, 2\pi \hbar c / k_B T_2)$ are valid. In this case, when emitting thermal photons, the particle can be considered as a point-like dipole with fluctuating dipole and magnetic moments $\mathbf{d}(t), \mathbf{m}(t)$. Material properties are taken into consideration through the frequency-dependent dielectric and (or) magnetic polarizabilities $\alpha_e(\omega)$, $\alpha_m(\omega)$ which are given in the particle rest frame $\Sigma''$.

Let the surface $\sigma$ encircles the particle at a large enough distance so that fluctuation electromagnetic field on the surface $\sigma$ can be considered as the wave field. According to the conventional form of the energy conservation law in the system within the volume V restricted by the external closed surface $\sigma$, we may write

$$-\frac{dW}{dt} = \oint_{\sigma} \mathbf{S} \cdot d\vec{\sigma} + \int_{V} \langle \mathbf{j} \cdot \mathbf{E} \rangle d^3 r \ ,$$

(1)

where

$$W = \int_{V} \frac{\langle \mathbf{E}^2 \rangle + \langle \mathbf{H}^2 \rangle}{8\pi} d^3 r$$

(2)

and

$$\mathbf{S} = \frac{c}{4\pi} \langle \mathbf{E} \times \mathbf{H} \rangle . \tag{3}$$

denote the energy of a fluctuating electromagnetic field in volume $V$ restricted by surface $\sigma$ and the Pointing vector of this field. The second term in (1) represents the Joule energy dissipation integral, and the angular brackets in (1)—(3) denote total quantum and statistical averaging. The wave-field on the surface $\sigma$ has the wave character. Within the quasistationary approximation used ($dW/dt = 0$) and from (1) one obtains the general expression for the intensity of radiation

$$I = \oint_{\sigma} \mathbf{S} \cdot d\vec{\sigma} = -\int_{V} \langle \mathbf{j} \cdot \mathbf{E} \rangle d^3 r \equiv I_1 - I_2 , \tag{4}$$

where $I_1 = I_1(T_1)$ is the net intensity of thermal radiation emitted by the particle in vacuum, and $I_2 = I_2(T_2)$ is the intensity of background radiation absorbed by the particle.

Our next step is to represent the right-hand side of (4) in a more convenient form. In the case of stationary electromagnetic fluctuations, using relativistic transformations for the density, of current and charge, electric field and volume in systems $\Sigma$ and $\Sigma'$, we obtain ($\beta = V/c$) [3]

$$\int_{V'} \langle \mathbf{j}' \cdot \mathbf{E}' \rangle d^3 r' = \frac{1}{1-\beta^2} \left( \int_{V} \langle \mathbf{j} \cdot \mathbf{E} \rangle d^3 r - F_x \cdot V \right) \equiv \frac{1}{1-\beta^2} \, dQ/dt , \tag{5}$$

$$dQ/dt = \langle \dot{\mathbf{d}} \cdot \mathbf{E} + \dot{\mathbf{m}} \cdot \mathbf{H} \rangle , \tag{6}$$

$$\mathbf{F} = \int \langle \rho \mathbf{E} \rangle d^3 r + \frac{1}{c} \int \langle \mathbf{j} \times \mathbf{H} \rangle d^3 r = \langle \nabla (\mathbf{d} \cdot \mathbf{E} + \mathbf{m} \cdot \mathbf{H}) \rangle . \tag{7}$$

In Eq. (5), $F_x$ is the projection of force $\mathbf{F}$ onto the velocity direction (only this component of force differs from zero), the points above the dipole moments in (6) denote the time derivative. Up to this point, formulas (1)-(7) coincide with those in [21], where the particle has no spin. In that case, the Joule dissipation integral (in the left-hand side of (5)) determines simultaneously the rate of particle heating $dQ'/dt'$ in $\Sigma'$. When the particle rotates with angular velocity $\Omega'$ with respect to $\Sigma'$, the work of fluctuating field is spent also on stopping this rotation. In this case, as shown in Appendix 1, one obtains

$$\int_V \langle \mathbf{j}' \cdot \mathbf{E}' \rangle d^3 r' = \langle \mathbf{d}' \cdot \mathbf{E}' \rangle = \dot{Q}' = \dot{Q}'' + \bar{M}' \bar{\Omega}',$$ (8)

where $\dot{Q}''$ is the heating rate of the particle in $\Sigma''$ and $\bar{M}'$ is the braking torque in $\Sigma'$. Due to nonrelativistic rotation of reference frame $\Sigma''$ relative to $\Sigma'$, we have $dt'' = dt' = dt(1-\beta^2)^{1/2}$, $\bar{\Omega}' = \bar{\Omega}$, and at $\bar{\Omega} = (\Omega, 0, 0)$ one obtains

$$\dot{Q}'' = C_0 dT_1 / dt' = C_0 (1-\beta^2)^{-1/2} dT_1 / dt,$$ (9)

$$\bar{M}' \bar{\Omega}' = M'_x \Omega = (1-\beta^2)^{-1/2} M_x \Omega.$$ (10)

where $C_0$ is the heat capacity of particle and the relativistic transformation of torque has been used. From (5) and (8)—(10) it follows

$$\frac{dT_1}{dt} = \frac{\dot{Q} - (1-\beta^2)^{1/2} M_x \Omega}{C_0 (1-\beta^2)^{1/2}},$$ (11)

where both $\dot{Q}$ and $M_x$ are given in reference frame $\Sigma$.

With allowance for (4) and (5) the intensity of radiation takes the form

$$I = I_1 - I_2 = -\left( \frac{dQ}{dt} + F_x V \right),$$ (12)

where $dQ/dt$ and $F_x$ are given by (6) and (7). It is worth noting that the relationship between $dQ'/dt'$ and $dQ/dt$ through (5) and (8) formally corresponds to the Planck formulation of relativistic thermodynamics assuming that $Q$ is associated with the amount of heat in reference frame $\Sigma$. Using (12), we can also rewrite Eq. (11) in another form which can be more convenient in certain cases

$$\frac{dT_1}{dt} = -\frac{I_1 - I_2 + F_x V + (1-\beta^2)^{1/2} M_x \Omega}{C_0 (1-\beta^2)^{1/2}}.$$ (13)

To complete this consideration, we write down an obvious equation for the braking torque of the particle in reference frame $\Sigma$

$$\mathbf{M} = \langle \mathbf{d} \times \mathbf{E} \rangle + \langle \mathbf{m} \times \mathbf{H} \rangle. \tag{14}$$

Equations (5)—(12) are the basis for our further calculations of the intensity of thermal radiation. The appearance of Eq. (12) is formally the same as in the case of a spinless particle [21], but all the quantities in the right-hand side depend on the angular velocity $\Omega$. For calculating the quantities $\dot{Q}, F_x$ and projections of torque $M_{x,z}$ we use our general method [3], representing the quantities in (6), (7), (14) as the products of spontaneous and induced components

$$\begin{cases} dQ/dt = \left\langle \dot{\mathbf{d}}^{sp} \mathbf{E}^{ind} + \dot{\mathbf{d}}^{ind} \mathbf{E}^{sp} + \dot{\mathbf{m}}^{sp} \mathbf{H}^{ind} + \dot{\mathbf{m}}^{ind} \mathbf{H}^{sp} \right\rangle \\ F_x = \left\langle \partial_x \left( \mathbf{d}^{sp} \mathbf{E}^{ind} + \mathbf{d}^{ind} \mathbf{E}^{sp} + \mathbf{m}^{sp} \mathbf{H}^{ind} + \mathbf{m}^{ind} \mathbf{H}^{sp} \right) \right\rangle \\ \mathbf{M} = \mathbf{d}^{sp} \times \mathbf{E}^{ind} + \mathbf{d}^{ind} \times \mathbf{E}^{sp} + \mathbf{m}^{sp} \times \mathbf{H}^{ind} + \mathbf{m}^{ind} \times \mathbf{H}^{sp} \end{cases} \tag{15}$$

All the quantities in the right-hand side of (15) have to be Fourier-transformed over the time and space variables $t, x, y\ z$. The points above the dipole moments indicate the time differentiation. The induced dipole moments $\mathbf{d}^{ind}, \mathbf{m}^{ind}$ have to be expressed through the fluctuating fields $\mathbf{E}^{sp}, \mathbf{H}^{sp}$ via linear integral relations. The induced fields $\mathbf{E}^{ind}, \mathbf{H}^{ind}$ have to be expressed through $\mathbf{d}^{sp}, \mathbf{m}^{sp}$ when solving the Maxwell equations containing the fluctuating currents induced by the dipole moments $\mathbf{d}^{sp}, \mathbf{m}^{sp}$. The arising correlators of the dipole moments and fields are calculated with the help of fluctuation-dissipation relations.

## 3. Intensity of radiation

In the case shown in Fig. 1a ($\bar{\Omega} = (\Omega, 0, 0)$), the general expressions for $F_x, \dot{Q}$ and $M_x$ were obtained in [19] and have the form

$$F_x = -\frac{\hbar \gamma}{4\pi c^4} \int_{-\infty}^{+\infty} d\omega \omega^4 \int_{-1}^{1} dxx \left\{ (1-x^2)(1-\beta^2)\alpha''(\omega_\beta) \cdot \left[ \coth \frac{\hbar \omega}{2k_B T_2} - \coth \frac{\hbar \omega_\beta}{2k_B T_1} \right] + \right.$$
$$\left. + \left[ (1+x^2)(1+\beta^2) + 4\beta x \right] \alpha''(\omega_\beta + \Omega) \cdot \left[ \coth \frac{\hbar \omega}{2k_B T_2} - \coth \frac{\hbar (\omega_\beta + \Omega)}{2k_B T_1} \right] \right\} \tag{16}$$

$$dQ/dt = \frac{\hbar\gamma}{4\pi c^3} \int\limits_{-\infty}^{+\infty} d\omega\,\omega^4 \int\limits_{-1}^{1} dx(1+\beta x)\left\{(1-x^2)(1-\beta^2)\alpha''(\omega_\beta)\cdot\left[\coth\frac{\hbar\omega}{2k_B T_2} - \coth\frac{\hbar\omega_\beta}{2k_B T_1}\right]+\right.$$
$$\left.+\left[(1+x^2)(1+\beta^2)+4\beta x\right]\alpha''(\omega_\beta+\Omega)\cdot\left[\coth\frac{\hbar\omega}{2k_B T_2} - \coth\frac{\hbar(\omega_\beta+\Omega)}{2k_B T_1}\right]\right\} \qquad (17)$$

$$M_x = -\frac{\hbar\gamma}{4\pi c^3}\int\limits_{-\infty}^{+\infty} d\omega\,\omega^3 \int\limits_{-1}^{1} dx\left[(1+x^2)(1+\beta^2)+4\beta x\right]\alpha''(\omega_\beta+\Omega)\cdot$$
$$\cdot\left[\coth\frac{\hbar\omega}{2k_B T_2} - \coth\frac{\hbar(\omega_\beta+\Omega)}{2k_B T_1}\right] \qquad (18)$$

In (16)—(18), we used notations $\gamma=(1-\beta^2)^{-1/2}$, $\omega_\beta=\gamma\omega(1+\beta x)$. We also did not specify the type of particle polarizability, assuming either electric $\alpha_e(\omega)$ or magnetic $\alpha_m(\omega)$ polarizability, or their sum; $\alpha''(\omega)$ denotes the imaginary part of $\alpha(\omega)$. Substituting (16), (17) into (12) yields

$$I = I_1 - I_2 = \frac{\hbar\gamma}{2\pi c^3}\int\limits_{0}^{+\infty} d\omega\,\omega^4 \int\limits_{-1}^{1} dx\cdot\left\{2(1-x^2)(1-\beta^2)\alpha''(\omega_\beta)\cdot\left[n_1(\omega_\beta)-n_2(\omega)\right]+\right.$$
$$\left.+\left[(1+x^2)(1+\beta^2)+4\beta x\right]\cdot\left(\begin{array}{l}\alpha''(\omega_\beta+\Omega)\cdot\left[n_1(\omega_\beta+\Omega)-n_2(\omega)\right]+\\ +\alpha''(\omega_\beta-\Omega)\cdot\left[n_1(\omega_\beta-\Omega)-n_2(\omega)\right]\end{array}\right)\right\} \qquad (19)$$

where $\coth(x/2)=1+\dfrac{2}{\exp(x)-1}$ and

$$n_i(x) = \frac{1}{\exp(x/\vartheta_i)-1},\ \vartheta_i=k_B T_i/\hbar,\ i=1,2\,. \qquad (20)$$

From (19), the intensity of radiation in $\Sigma$ takes the form

$$I_1 = \frac{\hbar\gamma}{2\pi c^3}\int\limits_{0}^{+\infty} d\omega\,\omega^4 \int\limits_{-1}^{1} dx\cdot\left\{\begin{array}{l}2(1-x^2)(1-\beta^2)\alpha''(\omega_\beta)n_1(\omega_\beta)+\left[(1+x^2)(1+\beta^2)+4\beta x\right]\cdot\\ \cdot\left[\alpha''(\omega_\beta+\Omega)n_1(\omega_\beta+\Omega)+\alpha''(\omega_\beta-\Omega)n_1(\omega_\beta-\Omega)\right]\end{array}\right\} \qquad (21)$$
.

In the case of motionless particle without spin, $V=0, \Omega=0$, Eq. (21) reduces to the well-known result which follows directly from the Kirchhoff law [27]

$$I_1 = \frac{4\hbar}{\pi c^3} \int\limits_0^\infty d\omega \, \omega^4 \alpha''(\omega) \frac{1}{\exp(\hbar\omega / k_B T_1) - 1}, \tag{22}$$

while at $\Omega = 0, V \neq 0$ we obtain [21]

$$I_1(T_1) = \frac{2\hbar\gamma}{\pi c^3} \int\limits_0^\infty d\omega \, \omega^4 \int\limits_{-1}^{1} dx (1 + \beta x)^2 \alpha''(\omega_\beta) \left[ \exp(\hbar\,\omega_\beta / k_B T_1) - 1 \right]^{-1}. \tag{23}$$

According to (21), the spectral-angular intensity distribution of thermal radiation is given by ($d\tilde{\Omega} = 2\pi \sin\theta \, d\theta$)

$$\frac{d^2 I}{d\omega \, d\tilde{\Omega}} = \frac{\gamma \hbar \omega^4}{4\pi^2 c^3} \cdot \left\{ \begin{array}{l} 2(1 - \beta^2)(1 - \cos^2\theta)\alpha''\big(\gamma\omega(1 - \beta\cos\theta)\big) n_1\big(\gamma\omega(1 - \beta\cos\theta)\big) + \\ + \big[(1 + \cos^2\theta)(1 + \beta^2) - 4\beta\cos\theta\big] \cdot \\ \left[ \begin{array}{l} \alpha''\big(\gamma\omega(1 - \beta\cos\theta) + \Omega\big) n_1\big(\gamma\omega(1 - \beta\cos\theta) + \Omega\big) + \\ + \alpha''\big(\gamma\omega(1 - \beta\cos\theta) - \Omega\big) n_1\big(\gamma\omega(1 - \beta\cos\theta) - \Omega\big) \end{array} \right] \end{array} \right\} \tag{24}$$

where $\theta$ is the angle between the particle velocity and the wave vector of radiation. From (24) one can see that the uniform motion and rotation of particle significantly should affect the spectral and angular distributions of thermal radiation. This is demonstrated in what follows.

## 4. Nonthermal radiation at $\Omega \neq 0$

A remarkable consequence of Eq. (21) is the fact that a rotating particle emits photons even at $T_1 = 0$. To prove this, we perform in (21) the limiting transition $T_1 \to 0$ by means of the relations

$$\lim_{T_1} n_1(\omega) = \frac{1}{2}\big(sign(\omega) - 1\big). \tag{25}$$

With allowance for (25) Eq. (21) takes the form

$$I_1 = \frac{\gamma \hbar}{2\pi c^3} \int\limits_{-1}^{+1} dx \int\limits_0^{\Omega\gamma^{-1}(1 + \beta x)^{-1}} d\omega \, \omega^4 \big[(1 + x^2)(1 + \beta^2) + 4\beta x\big] \alpha''\big(\Omega - \gamma\omega(1 + \beta x)\big) =$$
$$= \frac{4\hbar}{3\pi c^3} \int\limits_0^\Omega d\xi \, \xi^4 \alpha''(\Omega - \xi) \tag{26}$$

The result in the second line of (26) means that the total emitted radiation intensity does not depend on the particle velocity. At the same time, making use the transition $T_2 \to 0$ in (19), we obtain $I_2(0) = 0$, i. e. the particle does not absorb the external radiation. Coming back to the first line of Eq. (26), we can see that the distribution of the spectral-angular intensity of radiation depends on the velocity and takes the form ($d\tilde{\Omega} = 2\pi \sin\theta \, d\theta$)

$$\frac{d^2 I_1}{d\omega \, d\tilde{\Omega}} = \frac{\gamma \hbar \omega^4}{4\pi^2 c^3} \Theta(\Omega - \gamma\omega(1 - \beta\cos\theta))\left[(1 + \cos^2\theta)(1 + \beta^2) - 4\beta\cos\theta\right] \cdot$$
$$\cdot \alpha''(\Omega - \gamma\omega(1 - \beta\cos\theta)) \tag{27}$$

where $\Theta(x)$ is the unit Heaviside function. As follows from (27), the spontaneous radiation takes the frequency range

$$0 < \omega < \omega_{max} = \Omega\gamma^{-1}(1 - \beta\cos\theta)^{-1}. \tag{28}$$

The maximum frequency is emitted in the forward direction $\theta = 0$:

$$\omega_{max}(0) = \Omega\sqrt{\frac{1 + \beta}{1 - \beta}} \ . \tag{29}$$

In the ultrarelativistic case $\beta \to 1, \gamma \gg 1$ it follows $\omega_{max}(0) \approx 2\gamma\Omega \gg \Omega$. In the opposite direction ($\theta = \pi$) Eq. (28) yields $\omega_{max}(\pi) \approx \Omega/2\gamma \ll \Omega$. Therefore, the radiation spectrum is strongly anisotropic. At low velocities $\beta \ll 1$, correspondingly, $\omega_{max}(0) = \Omega(1 + \beta)$, $\omega_{max}(\pi) = \Omega(1 - \beta)$. At $\beta = 0$, the spectrum (27) takes the simplest form, though a small angular dependence still takes place :

$$\frac{d^2 I_1}{d\omega \, d\Omega} = \frac{\hbar \omega^4}{4\pi^2 c^3} \Theta(\Omega - \omega)\left(1 + \cos^2\theta\right)\alpha''(\Omega - \omega) \tag{30}$$

These spectral properties can be used for the experimental verification. In Sec. 5 the spectral properties of this nonthermal radiation will be analyzed further.

It is worthwhile also to consider the characteristics of particle dynamics and heating. In particular, the particle thermal state is crucially important since the initial condition $T_1 = 0$ is violated with time. By performing transitions $T_1 = T_2 \to 0$ in Eqs. (16), (18) one obtains

$$F_x = -\frac{4\hbar\beta}{3\pi c^4}\int_0^\Omega d\xi\xi^4\alpha''(\Omega-\xi) \tag{31}$$

$$M_x = -\frac{4\hbar}{3\pi c^3\gamma}\int_0^\Omega d\xi\xi^3\alpha''(\Omega-\xi) \tag{32}$$

Furthermore, substituting (26), (31) and (32) into (13) yields

$$C_0\frac{dT_1}{dt} = \frac{4\hbar}{3\pi c^3\gamma}\int_0^\Omega d\xi\xi^3(\Omega-\xi)\alpha''(\Omega-\xi) \tag{33}$$

It is worthwhile to emphasize once again that temperature $T_1$ corresponds to the rest frame of particle, and time $t$ corresponds to the rest frame of vacuum. According to Eq. (31), a rotating particle in a vacuum experiences frictional drag, i. e. cold vacuum becomes viscous. As follows from (32), the rotation is also decelerated, while from (33) one can see that the particle is heated. This means that instead of Eqs. (27), (31), (32) one should use the more general formulas (24), (16), (18), depending on $T_1$. In this case, the nonthermal radiation will be mixed with thermal component, but the former one does not disappear. Moreover, at a large enough characteristic time of heating (see Sec. 5), nonthermal spectrum (27) will be observed for rather long time.

In the case of particle rotation around the $z$ axis (Fig. 1b), the corresponding basic formulas are given in Appendix B.

## 5. Special case: conductive particle with the Drude-like dielectric properties

In further analysis, we discuss the case where the particle polarization corresponds to the low-frequency limit of the Drude dielectric permittivity $\varepsilon(\omega) \approx i\cdot 4\pi\sigma_0/\omega$, where $\sigma_0$ is the static conductivity. Respectively, the electric and magnetic polarizabilities of a spherical particle with radius $R$ are given by [28]

$$\alpha_e''(\omega) = 3R^3\omega/4\pi\sigma_0 \tag{34}$$

$$\alpha_m''(\omega) = -\frac{3Rc^2}{8\pi^2\sigma_0\omega}\chi(x), \quad x \equiv 2R\cdot(2\pi\sigma_0\omega)^{1/2}/c \tag{35}$$

$$\chi(x) = 1 - \frac{x}{2}\frac{\sinh x + \sin x}{\cosh x - \cos x} \tag{36}$$

For $R << c/(2\pi\sigma_0\omega)^{1/2}$, from (35) and (36) it follows

$$\alpha_m'' \approx \frac{2\pi R^5\sigma_0}{15c^2}\omega, \tag{37}$$

and in this case $\alpha_m''$ differs from $\alpha_e''$ by only numerical factor $\frac{8\pi^2}{45}\left(\frac{R\sigma_0}{c}\right)^2$.

### a) nonthermal radiation at $T_1 = T_2 = 0$

First, let us compare the time scales of particle rotation and heating, assuming $\beta << c$. Using (32) and the dynamics equation

$$I\,d\Omega/dt = M_x, \tag{38}$$

where $I = (8\pi/15)R^5\rho$ is the inertia moment of spherical particle with density $\rho$, one obtains

$$\Omega = \Omega_0\left(1 + \frac{3\hbar\Omega_0^4 t}{8\pi^3\rho R^2\sigma_0 c^3}\right)^{-1/4} \tag{39}$$

where $\Omega_0$ is the initial angular velocity. From (39) and assuming that $\sigma_0 = const$, the characteristic decay time $t_{1/2}$ corresponding to a decrease of $\Omega$ by two times is given by

$$t_{1/2} = \frac{40\pi^3\rho R^2\sigma_0 c^3}{\hbar\Omega_0^4} \tag{40}$$

On the other hand, with allowance for the fact that at low temperatures the heat capacity of a metallic particle can be written as $C_0 = \frac{4\pi}{3}\rho R^3\cdot aT$ ($a$ is a numerical coefficient), and assuming $\Omega \approx const = \Omega_0$, from (33) one obtains the time needed to increase the particle temperature from zero to $T = T_1$:

$$\tau \approx \frac{40\pi^3 \rho \,\sigma_0 c^3 a T_1^{\,2}}{\hbar \,\Omega_0^{\,6}}.$$ (41)

From (40) and (41) it follows

$$t_{1/2}/\tau = \left(\frac{R\Omega_0}{T_1}\right)^2 \frac{1}{a}$$ (42)

For example, at $\Omega_0 = 10^{12}\, s^{-1}, R = 2nm, T_1 = 1K$ and $a = 60\, erg/g \cdot K^2$ [29] (gold) we obtain $t_{1/2}/\tau = 10^9$, whereas the value of $\tau$ is estimated to be more than 100 years, since for gold at $\sigma_0 \gg 10^{17} s^{-1}$ if $T_1 \to 0$. This means that the condition of an approximate constancy of the angular velocity can be fulfilled during a rather long time, while the spectrum of radiation will be described by Eq. (27).

As we have shown in [21], the time of stopping is always much less than the time of heating. Therefore, during the time of heating the impact of particle linear deceleration on the nonthermal radiation spectrum (27) is negligible.

Next we discuss the radiation spectrum in more detail. By inserting (34) $(\alpha'' \sim \omega)$ into (27) and integrating over the frequencies one obtains

$$\frac{dI_1}{d\tilde{\Omega}} = \frac{\hbar\,\Omega^3}{160\,\pi^3\sigma_0}\left(\frac{R\,\Omega}{c}\right)^3 \frac{\left[(1+\cos^2\theta)(1+\beta^2) - 4\beta\cos\theta\right]}{\gamma^4(1-\beta\cos\theta)^5}$$ (43)

One can see from (43) that the angular spectrum is strongly anisotropic and the maximum intensity in the forward direction is given by

$$\left(\frac{dI_1}{d\tilde{\Omega}}\right)_{\theta=0} = \frac{\hbar\,\Omega^3}{80\,\pi^3\sigma_0}\left(\frac{R\,\Omega}{c}\right)^3 (1+\beta)^3\gamma^2$$ (44)

The spectral distribution of radiation is obtained by using (34) and integrating (27) over the solid angle $\tilde{\Omega}$

$$\frac{dI_1}{d\omega} = \frac{3}{8\pi^2} \frac{\hbar\Omega^2}{\beta\,\sigma_0} \left(\frac{\Omega R}{c}\right)^3 S(w,\beta,\gamma) w^3, \, w \equiv \omega/\Omega \tag{45}$$

where the function $S(w,\beta,\gamma)$ is given in Appendix C together with the resulting formula for configuration $\bar{\Omega} = (0,0,\Omega)$ (Fig. 1b). The total intensity of radiation is obtained when integrating (43) or (45) and has the form

$$I_1 = \frac{\hbar\Omega^3}{30\pi^2\sigma_0} \left(\frac{\Omega R}{c}\right)^3 \tag{46}$$

A remarkable fact is that this nonthermal intensity does not depend on the particle velocity, in contrast to the angular and frequency distributions of the intensity. This is a consequence of (26) and does not depend on the functional form of polarizability.

Figure 2 shows the angular intensity distributions (43) normalized to the maximum intensity (44), depending on the velocity factor $\beta$. Also shown is the distribution in configuration Fig. 1b, Thick solid and dashed lines and thin (solid and dashed) lines correspond to configurations (a) and (b) in Fig. 1, respectively. Figure 3 shows the spectral-frequency distributions normalized to the factor $\frac{3}{8\pi^2} \frac{\hbar\Omega^2}{\sigma_0} \left(\frac{\Omega R}{c}\right)^3$ calculated by Eq. (45), depending on the velocity factor $\beta$. As in Fig. 2, the thick and thin lines correspond to configurations (a) and (b) in Fig.1. One can see that at small velocities (Fig. 2a) the angular spectra considerably depend on configuration, whereas with increasing velocity (Fig. 2b) this dependence disappears and the spectra are strongly concentrated in the forward direction. With increasing relativistic factor $\gamma$ the frequency spectrum becomes broader up to the maximum possible frequency $\omega_{max} = 2\gamma\Omega$.

**b) finite temperature conditions and radiation**

At $T_1 > 0$ and $\beta \neq 0$, the radiation exists even at $\Omega = 0$ [21]. In the general case $\beta \neq 0, \Omega \neq 0$, the characteristics of spectrum depend on $\beta$ and $\Omega$ (see Eq. (24)). For the particular case of particle polarizability (34), using Eqs. (16)—(19), (21) and performing integrations explicitly one obtains ($\vartheta_i = k_B T_i / \hbar, i = 1,2$)

$$F_x = -\frac{8\pi^4}{21} \frac{\hbar R^3}{c^4 \sigma_0} \beta \left[ \frac{(1+\beta^2/5)}{(1-\beta^2)} \vartheta_2^{\,6} + \vartheta_1^{\,6} \psi_1(\Omega/\vartheta_1) \right], \tag{47}$$

$$\dot{Q} = \frac{8\pi^4}{21} \frac{\hbar R^3}{c^3 \sigma_0} \left[ \frac{(1+2\beta^2 + \beta^4/5)}{(1-\beta^2)} \vartheta_2{}^6 - (1-\beta^2)\vartheta_1{}^6 \psi_1(\Omega/\vartheta_1) \right],$$ (48)

$$I = \frac{8\pi^4}{21} \frac{\hbar R^3}{c^3 \sigma_0} \left[ \vartheta_1{}^6 \psi_1(\Omega/\vartheta_1) - \frac{(1+\beta^2)}{(1-\beta^2)} \vartheta_2{}^6 \right],$$ (49)

$$I_1 = \frac{8\pi^4}{21} \frac{\hbar R^3}{c^3 \sigma_0} \vartheta_1{}^6 \psi_1(\Omega/\vartheta_1),$$ (50)

$$M_x = -\frac{2\pi^2}{15} \frac{\hbar R^3 \gamma \Omega}{c^3 \sigma_0} \left[ (1+\beta^2)\vartheta_2{}^4 + \vartheta_1{}^4(1-\beta^2)\psi_2(\Omega/\vartheta_1) \right],$$ (51)

$$\psi_1(x) = 1 + \frac{21}{10\pi^2} x^2 + \frac{7}{8\pi^4} x^4 + \frac{7}{80\pi^6} x^6 ,$$ (52)

$$\psi_2(x) = 3 + \frac{5}{2\pi^2} x^2 + \frac{3}{8\pi^4} x^4 .$$ (53)

Using (11), (48) and (51) yields

$$dT_1/dt = \frac{2\pi^3}{7} \frac{\hbar}{c^3 \sigma_0 \rho C_s} \cdot$$

$$\cdot \left[ \begin{array}{l} \dfrac{(1+2\beta^2+\beta^4/5)}{(1-\beta^2)^{3/2}} \vartheta_2{}^6 - \vartheta_1{}^6 \sqrt{1-\beta^2}\, \psi_1(\Omega/\vartheta_1) + \\[2mm] + \dfrac{7\Omega^2}{20\pi^2} \left( \dfrac{1+\beta^2}{\sqrt{1-\beta^2}} \vartheta_2{}^4 + \sqrt{1-\beta^2}\, \vartheta_1{}^4 \psi_2(\Omega/\vartheta_1) \right) \end{array} \right],$$ (54)

where $C_s = 3C_0/4\pi\rho R^3$. Moreover, using the dynamics equation $mcd\beta/dt = (1-\beta^2)^{3/2} F_x$ and (47) yields

$$d\beta/dt = -\frac{2\pi^3}{7} \frac{\hbar}{c^5 \rho \sigma_0} \beta(1-\beta^2)^{3/2} \left[ \frac{(1+\beta^2/5)}{(1-\beta^2)} \vartheta_2{}^6 + \vartheta_1{}^6 \psi_1(\Omega/\vartheta_1) \right].$$ (55)

Interestingly, formulas (47)—(50) and (55) prove to be valid also in configuration (b) (Fig. 1b). However, the formulas for $M_z$ and $dT_1/dt$ in this case are different and take the form

$$M_z = -\frac{2\pi^2}{15} \frac{\hbar R^3 \Omega}{c^3 \sigma_0} \left[ \vartheta_2{}^4 + \vartheta_1{}^4 \psi_2(\Omega/\vartheta_1) \right],$$ (56)

$$dT_1 / dt = \frac{2\pi^3}{7} \frac{\hbar}{c^3 \sigma_0 \rho C_s} \cdot$$

$$\cdot \left[ \frac{(1 + 2\beta^2 + \beta^4 / 5)}{(1 - \beta^2)^{3/2}} \vartheta_2{}^6 - \vartheta_1{}^6 \sqrt{1 - \beta^2} \psi_1(\Omega / \vartheta_1) + \frac{7\Omega^2}{20\pi^2} \frac{1}{\sqrt{1 - \beta^2}} \left( \vartheta_2{}^4 + \vartheta_1{}^4 \psi_2(\Omega / \vartheta_1) \right) \right] \cdot \qquad (57)$$

Formulas (45)—(57) provide the basis to control the dynamical/thermal state of particles and their integral radiation.

As was shown in [21] in the case of a spinless particle ($\Omega = 0$) and $T_2 = const$ (when the thermal state of vacuum is not changed), the characteristic time scale of particle cooling/heating $\tau_Q$ is many orders of magnitude lower than the time scale $\tau_\beta$ of particle stopping. Owing to this, the particle rather quickly reaches a quasi-equilibrium thermal state (it its own rest frame) with the temperature $T_e = \kappa T_2$ where the factor $\kappa$ is an increasing function of $\beta$. At $\Omega > 0$ this picture is not changed qualitatively. Moreover, simple analysis of Eqs. (51), (54) and (55) shows that the characteristic time scale of rotational deceleration $\tau_M$ is much longer than the thermal time scale $\tau_Q$ ( similarly to the "cold" case), but $\tau_M \ll \tau_\beta$, i. e. $\tau_Q \ll \tau_M \ll \tau_\beta$.

Thus it follows that in the steady-state thermal regime one obtains $T_1 \rightarrow T_e = \kappa(\beta, \Omega) T_2$, and the particle spectrum will be determined by its effective temperature $T_e$ for a long time of angular and translational deceleration. The function $\kappa(\beta, \Omega)$ can be determined using the stationarity condition $dT_1 / dt = 0$ and Eqs. (54), (57). For example, at $\Omega = 0$ from (54) it follows

$$T_e = T_2 \cdot \left( \frac{1 + 2\beta^2 + \beta^4 / 5}{(1 - \beta^2)^2} \right)^{1/6}, \qquad (58)$$

whereas the corresponding intensity of radiation (by inserting (58) into (50)) is given by

$$I_1 = \frac{8\pi^4}{21} \frac{\hbar R^3}{c^3 \sigma_0} \left( \frac{k_B T_2}{\hbar} \right)^6 \frac{(1 + 2\beta^2 + \beta^4 / 5)}{(1 - \beta^2)^2} \qquad (59)$$

Comparing (59) with (46) we can see that the "steady-state" thermal radiation is expected to be much stronger than the nonthermal radiation, if $k_B T_2 / \Omega \cdot \hbar > 1$, and its magnitude increases further with increasing relativistic factor as $\sim \gamma^4$.

Another simple case corresponds to a motionless particle with spin ($\Omega > 0$ and $\beta = 0$) and a cold vacuum, $T_2 = 0$. In this case, using the stationarity condition $dT_1 / dt = 0$ and (54) one obtains $\vartheta_1 = 5.17\Omega$, i. e. the effective particle temperature is determined by the angular velocity. According to Eq. (50), the intensity of radiation takes the form

$$I_1 = 8.35 \cdot 10^6 \, \hbar R^3 \Omega^6 / c^3 \sigma_0 \qquad (60)$$

This intensity is by $\sim 3 \cdot 10^9$ times higher than (46). For example, assuming $\Omega = 10^{12} \, s^{-1}$, for a gold nanoparticle ($\sigma_0 \approx 10^{17} \, s^{-1}$) with a radius of $2 nm$, from (60) one obtains $I_1 \approx 3 \cdot 10^{-24} \, W$.

In the general case $V \neq 0, \Omega \neq 0, T_2 > 0$, the dependence $\kappa(\beta, \Omega)$ in the equation $T_e = \kappa(\beta, \Omega) T_2$ is calculated numerically. Figure 4 and Fig. 5 shows the calculated steady-state ratio $T_e / T_2$ and the normalized radiation intensity $\bar{I} = I_1 / I_0$ for a gold nanoparticle, depending on $\Omega / \vartheta_2$ and $\beta$. The normalization factor is $I_0 = \dfrac{8\pi^4}{21} \dfrac{\hbar R^3}{c^3 \sigma_0} \left( \dfrac{k_B T_2}{\hbar} \right)^6$. One can see that the impact of angular rotation velocity is rather small at $\beta \leq 0.9$ and $\Omega \hbar / k_B T_2 < 5$ and increases greatly at $\Omega \hbar / k_B T_2 > 5$ and $\beta > 0.9$ One can also note a universal character of $T_e / T_2$ and the normalized intensity at a high angular velocity $\Omega$ (irrespectively of $\beta$). A more detailed analysis of the frequency-angle spectral characteristics of thermal radiation requires special consideration.

## 7. Conclusions

We have formulated a theory, that describes the nonthermal and thermal radiation of a relativistic polarizable particle with spin moving in a vacuum background of arbitrary temperature. We have established that the particle emits the long-wavelength photons ($\omega R / c \ll 1$) even at zero temperature of particle and vacuum. The condition of nonthermal radiation meets the condition of Zel'dovich superradiance. This situation differs principally from its counterpart at $\Omega = 0, V = 0, T_1 = T_2 = 0$, when a vacuum background has no effect on the uniform motion of particle, in accordance with the principle of relativity. The presence of rotation changes the situation radically and allows one to speak about rotation relative to the vacuum environment. This implies a physical non-equivalence of inertial and non-inertial frames of reference. In particular, this non-equivalence also leads to finite viscosity of vacuum (Eq. (31)). Despite its minuteness, the observation of nonthermal radiation can be of great importance when studying the fundamental properties of vacuum. A striking feature is that the integral intensity of nonthermal radiation is independent of the particle velocity and the mutual orientation of spin and velocity vector. When the particle or/and vacuum background have finite temperatures, the thermal state of particle is stabilized much more faster than the characteristic time of slowing down, and the quasi-steady-state temperature of particle is determined by the values of angular/linear velocities and the background temperature. The state of equilibrium is reached irrespectively of the initial temperatures of particle and vacuum. At relativistic velocities, the particle temperature increases considerably and may reach the melting point. The intensity of radiation increases significantly and is concentrated in the velocity direction. The direction of particle spin relative to the velocity has a significant impact on the frequency-angular spectral distribution of radiation and the integral intensity.

These results can be of interest in creating new types of the directional microwave radiation, in studying particle motion in cavities and in astrophysics. Astrophysical applications involve the observations of microwave cosmic radiation from spacecrafts when investigating the gravitational compression of gaseous and dust clouds, and accretion of massive cosmic objects. The directional effect of thermal radiation of moving particles can probably influence the observed anisotropy of the primary 2.7 K blackbody radiation.

## Appendix A

Consider the transformation of product $\mathbf{d}' \cdot \mathbf{E}'$ (where $\mathbf{d}' = \partial \mathbf{d}'/\partial \tau$) when passing from the reference frame $\Sigma'$ to $\Sigma''$ (Fig. 1). The values $\mathbf{d}'(\tau)$, $\mathbf{E}'(\tau)$ and $\mathbf{d}''(\tau)$, $\mathbf{E}''(\tau)$ are considered as the random functions of the own time $\tau$ in $\Sigma'$ and $\Sigma''$, which is the same in both these systems since the rotation of $\Sigma''$ is assumed to be nonrelativistic. When rotating around the $x'$ axis with the angular velocity $\Omega$ ($\Omega$ is given in $\Sigma'$), we obtain

$$d_x'(\tau) = d_x''(\tau)$$
$$d_y'(\tau) = d_y''(\tau)\cos\Omega\tau - d_z''(\tau)\sin\Omega\tau \qquad\qquad (A1)$$
$$d_z'(\tau) = d_y''(\tau)\sin\Omega\tau + d_z''(\tau)\cos\Omega\tau$$

$$E_x'(\tau) = E_x''(\tau)$$
$$E_y'(\tau) = E_y''(\tau)\cos\Omega\tau - E_z''(\tau)\sin\Omega\tau \qquad\qquad (A2)$$
$$E_z''(\tau) = E_y''(\tau)\sin\Omega\tau + E_z''(\tau)\cos\Omega\tau$$

$$\mathbf{d}' \cdot \mathbf{E}' = d_x'E_x' + d_y'E_y' + d_z'E_z' \qquad\qquad (A3)$$

Substituting (A1),(A2) into (A3) yields

$$\mathbf{d}' \cdot \mathbf{E}' = \mathbf{d}'' \cdot \mathbf{E}'' + (d_y''E_z'' - d_z''E_y'')\Omega = \mathbf{d}'' \cdot \mathbf{E}'' + (\mathbf{d}'' \times \mathbf{E}'')_x \cdot \Omega \qquad\qquad (A4)$$

With allowance for the identity of torque $M_x = (\mathbf{d}'' \times \mathbf{E}'')_x$ in $\Sigma'$ and $\Sigma''$, we obtain from (A4)

$$\mathbf{d}' \cdot \mathbf{E}' = \mathbf{d}'' \cdot \mathbf{E}'' + M_x' \cdot \Omega \qquad (A5)$$

Using statistical averaging of (A5) with allowance for $\langle \mathbf{d}'' \cdot \mathbf{E}'' \rangle = \dot{Q}''$, where $\dot{Q}''$ is the rate of the internal energy change of spinning particle (heating rate), we obtain

$$\int\limits_V \langle \mathbf{j}' \cdot \mathbf{E}' \rangle d^3 r' = \langle \mathbf{d}' \cdot \mathbf{E}' \rangle = \dot{Q}'' + M_x' \Omega' = \dot{Q}'' + M_x' \Omega \qquad (A6)$$

Equation (A6) is easily generalized in the case of arbitrary angular velocity $\bar{\Omega} = (\Omega_x, \Omega_y, \Omega_z)$ of the system $\Sigma''$ relative to $\Sigma'$. In this case we obtain

$$\int\limits_V \langle \mathbf{j}' \cdot \mathbf{E}' \rangle d^3 r' = \langle \mathbf{d}' \cdot \mathbf{E}' \rangle = \dot{Q}'' + \bar{M}' \bar{\Omega}' \qquad (A7)$$

The same calculations can be performed for the contribution of magnetic moment $\mathbf{m}'$. The resultant Joule dissipation integral with allowance for the electric and magnetic polarization contributions takes the form (cf. with (A7))

$$\int\limits_V \langle \mathbf{j}' \cdot \mathbf{E}' \rangle d^3 r' = \langle \mathbf{d}' \cdot \mathbf{E}' + \dot{\mathbf{m}}' \cdot \mathbf{H} \rangle = \dot{Q}'' + \bar{M}' \bar{\Omega}' \qquad (A8)$$

Unlike (A7), the quantities of $\dot{Q}''$ and $\bar{M}'$ in (A8) involve both $\mathbf{d}'$ and $\mathbf{m}'$.

## Appendix B

In configuration $\Omega \perp x$ ($\Omega \| z$), the expressions for $F_x, \dot{Q}$ and $M_x$ are given by [19] :

$$F_x = -\frac{\hbar \gamma}{8\pi c^4} \int\limits_{-\infty}^{+\infty} d\omega \, \omega^4 \int\limits_{-1}^{1} dx \, x \cdot$$

$$\cdot \left\{ \begin{array}{l} \left[ (1+x^2)(1+\beta^2) + 2(1-\beta^2)(1-x^2) + 4\beta x \right] \alpha''(\omega_\beta + \Omega) \left[ \coth \dfrac{\hbar \omega}{2k_B T_2} - \coth \dfrac{\hbar(\omega_\beta + \Omega)}{2k_B T_1} \right] + \\[3mm] + \left[ (1+x^2)(1+\beta^2) + 4\beta x \right] \alpha''(\omega_\beta) \cdot \left[ \coth \dfrac{\hbar \omega}{2k_B T_2} - \coth \dfrac{\hbar \omega_\beta}{2k_B T_1} \right] \end{array} \right\} \qquad (B1)$$

$$\dot{Q} = \frac{\hbar \gamma}{8\pi c^3} \int\limits_{-\infty}^{+\infty} d\omega \, \omega^4 \int\limits_{-1}^{1} dx \, (1+\beta x) \cdot$$

$$\cdot \left\{ \begin{array}{l} \left[ (1+x^2)(1+\beta^2) + 2(1-\beta^2)(1-x^2) + 4\beta x \right] \alpha''(\omega_\beta + \Omega) \cdot \left[ \coth \dfrac{\hbar \omega}{2k_B T_2} - \coth \dfrac{\hbar(\omega_\beta + \Omega)}{2k_B T_1} \right] + \\[3mm] + \left[ (1+x^2)(1+\beta^2) + 4\beta x \right] \alpha''(\omega_\beta) \cdot \left[ \coth \dfrac{\hbar \omega}{2k_B T_2} - \coth \dfrac{\hbar \omega_\beta}{2k_B T_1} \right] \end{array} \right\} \qquad (B2)$$

$$M_z = -\frac{\hbar}{8\pi c^3} \int\limits_{-\infty}^{+\infty} d\omega\, \omega^3 \int\limits_{-1}^{1} dx \left(3 - x^2 + 2\beta x\right) \alpha''\left(\omega_\beta + \Omega\right) \cdot \left[\coth\frac{\hbar\omega}{2k_B T_2} - \coth\frac{\hbar(\omega_\beta + \Omega)}{2k_B T_1}\right] \tag{B3}$$

Using (B1)—(B3) yields

$$I = I_1 - I_2 = \frac{\hbar\gamma}{4\pi c^3} \int\limits_{0}^{+\infty} d\omega\, \omega^4 \int\limits_{-1}^{1} dx \cdot \left\{ 2\left[(1+x^2)(1+\beta^2) + 4\beta x\right]\alpha''(\omega_\beta) \cdot \left[n_1(\omega_\beta) - n_2(\omega)\right] + \right.$$
$$\left. + \left[(1+x^2)(1+\beta^2) + 2(1-x^2)(1-\beta^2) + 4\beta x\right] \cdot \left[\begin{array}{l} \alpha''\left(\omega_\beta + \Omega\right) \cdot \left[n_1(\omega_\beta + \Omega) - n_2(\omega)\right] + \\ \alpha''\left(\omega_\beta - \Omega\right) \cdot \left[n_1(\omega_\beta - \Omega) - n_2(\omega)\right] \end{array}\right] \right\} \tag{B4}$$

$$I_1 = \frac{\hbar\gamma}{4\pi c^3} \int\limits_{0}^{+\infty} d\omega\, \omega^4 \int\limits_{-1}^{1} dx \cdot \left\{ 2\left[(1+x^2)(1+\beta^2) + 4\beta x\right]\alpha''(\omega_\beta) \cdot n_1(\omega_\beta) + \right.$$
$$\left. + \left[(1+x^2)(1+\beta^2) + 2(1-x^2)(1-\beta^2) + 4\beta x\right] \cdot \left[\begin{array}{l} \alpha''\left(\omega_\beta + \Omega\right) n_1(\omega_\beta + \Omega) + \\ + \alpha''\left(\omega_\beta - \Omega\right) n_1(\omega_\beta - \Omega) \end{array}\right] \right\} \tag{B5}$$

$$\frac{d^2 I}{d\omega\, d\tilde{\Omega}} = \frac{\gamma\hbar\omega^4}{8\pi^2 c^3} \cdot \left\{ \begin{array}{l} 2\left[(1+\cos^2\theta)(1+\beta^2) - 4\beta\cos\theta\right]\alpha''(\gamma\omega(1-\beta\cos\theta))n_1\left(\gamma\omega(1-\beta\cos\theta)\right) + \\ + \left[(1+\cos^2\theta)(1+\beta^2) + 2(1-\cos^2\theta)(1-\beta^2) - 4\beta\cos\theta\right] \cdot \\ \cdot \left[\begin{array}{l} \alpha''\left(\gamma\omega(1-\beta\cos\theta) + \Omega\right)n_1\left(\gamma\omega(1-\beta\cos\theta) + \Omega\right) + \\ + \alpha''\left(\gamma\omega(1-\beta\cos\theta) - \Omega\right)n_1\left(\gamma\omega(1-\beta\cos\theta) - \Omega\right) \end{array}\right] \end{array} \right\} \tag{B6}$$

At $T_1 \to 0$ we obtain

$$I_1 = \frac{\gamma\hbar}{4\pi c^3} \int\limits_{-1}^{+1} dx \int\limits_{0}^{\Omega\omega/\omega_\beta} d\omega\, \omega^4 \left[(1+x^2)(1+\beta^2) + 2(1-\beta^2)(1-x^2) + 4\beta x\right]\alpha''\left(\Omega - \omega_\beta\right) =$$
$$= \frac{3\hbar}{4\pi c^3} \int\limits_{0}^{\Omega} d\xi\, \xi^3 \alpha''(\Omega - \xi) \tag{B7}$$

$$\frac{d^2 I_1}{d\omega\, d\tilde{\Omega}} = \frac{\gamma\hbar\omega^4}{8\pi^2 c^3} \Theta\left(\Omega - \gamma\omega(1-\beta\cos\theta)\right) \cdot$$
$$\cdot \left[(1+\cos^2\theta)(1+\beta^2) + 2(1-\beta^2)(1-\cos^2\theta) - 4\beta\cos\theta\right]\alpha''\left(\Omega - \gamma\omega(1-\beta\cos\theta)\right) \tag{B8}$$

It should be noted that due to the lack of the axial symmetry (in this case), the spectral-angular intensity distributions in (B6) and (B8) are averaged over the azimuthal angle $\varphi$.

# APPENDIX C

Formula (45) is obtained when substituting (34) in (27), introducing a new variable $\Omega - \gamma\,\omega(1 - \beta\cos\theta) \equiv y$ and integrating by parts. The resultant expression has the form

$$
\begin{aligned}
S(w,\beta,\gamma) = & f_1(\gamma\,w,\beta,1-\gamma\,w(1-\beta))\cdot\Theta(1-\gamma\,w(1-\beta)) - \\
& - f_1(\gamma\,w,\beta,1-\gamma\,w(1+\beta))\cdot\Theta(1-\gamma\,w(1+\beta))
\end{aligned}
\tag{C1}
$$

$$
f_1(a,b,x) = \frac{(1+b^2)}{4a^2b^2}x^4 + \frac{2}{3}\left[(1+b^2)\frac{(a-1)}{a^2b^2}-\frac{2}{a}\right]x^3 + \frac{1}{2}\left[(1+b^2)\left[1+\frac{(a-1)^2}{a^2b^2}\right]-\frac{4(a-1)}{a}\right]x^2
\tag{C2}
$$

In configuration (b) (Fig. 1b), function (C2) is replaced by

$$
\begin{aligned}
f_2(a,b,x) = & \frac{1}{4}\frac{(3b^2-1)}{a^2b^2}x^4 + \frac{2}{3}\left[\frac{(a-1)(3b^2-1)}{a^2b^2}-\frac{2}{a}\right]x^3 - \\
& - \frac{1}{2}\left[\frac{a^2b^4+6ab^2+a^2+1-6a^2b^2-3b^2-2a}{a^2b^2}+\frac{4(a-1)}{a}\right]x^2
\end{aligned}
\tag{C3}
$$

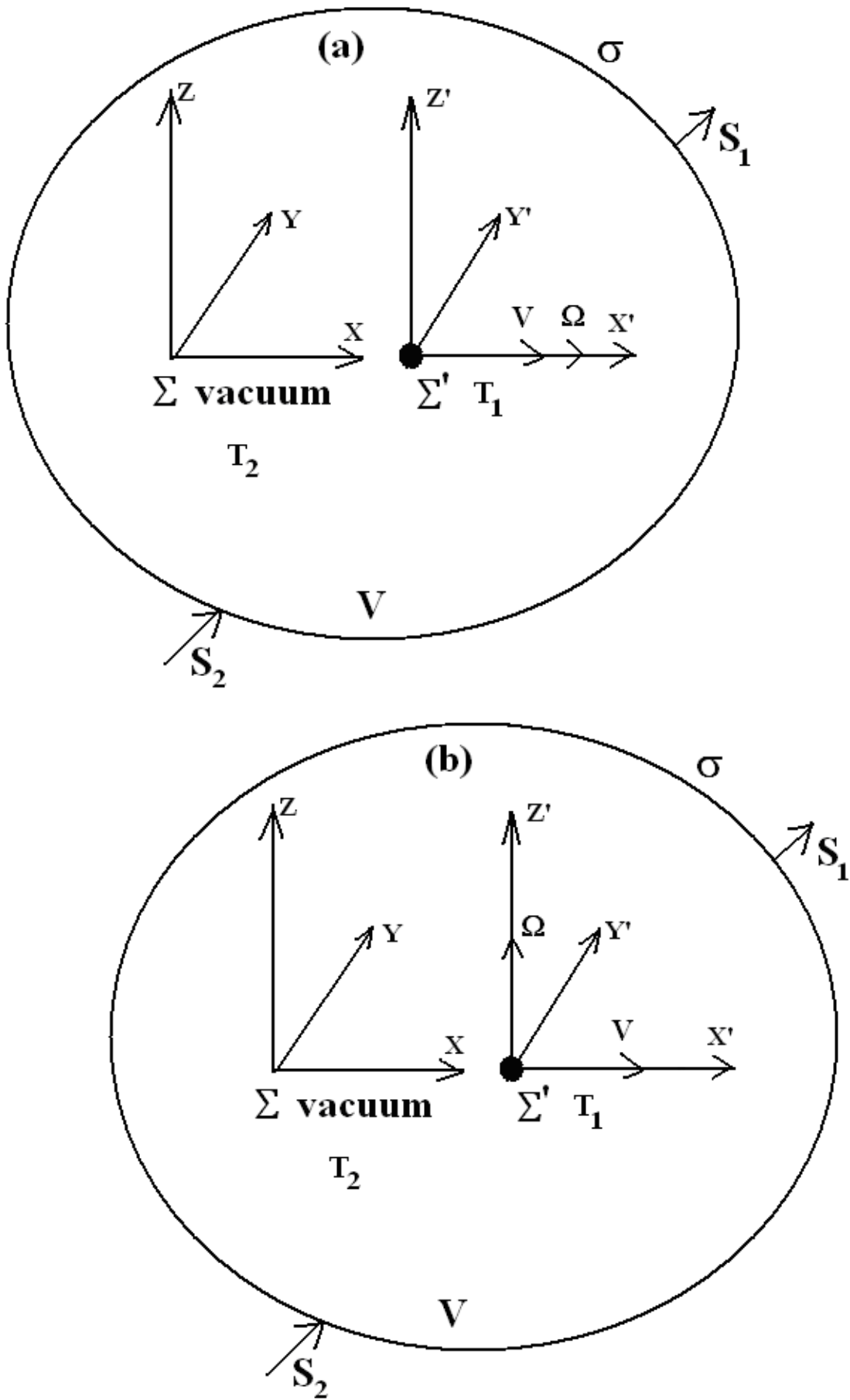

Fig.1. Configurations (a) and (b) of the system with different mutual orientation of the vectors of angular and linear velocities.

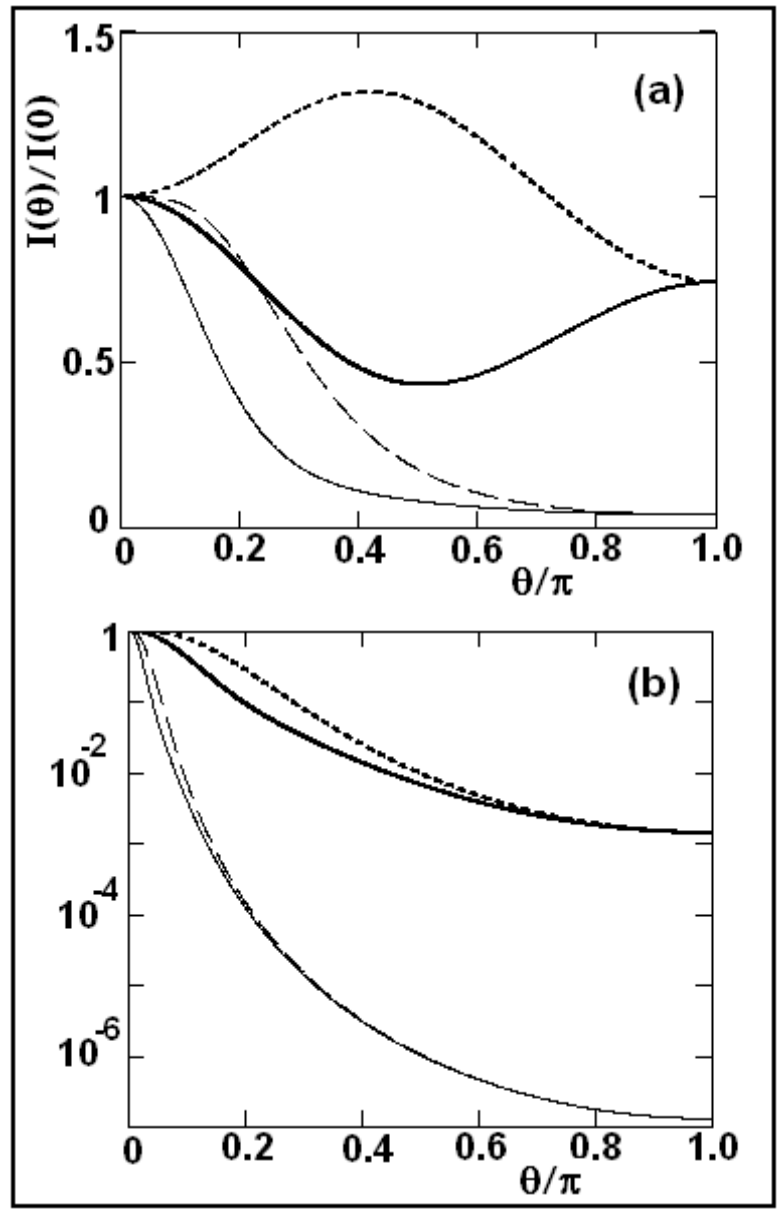

Fig. 2 Normalized angular spectrum of nonthermal radiation, depending on $\beta$ . Solid lines correspond to configuration $\vec{\Omega} = (0,0,\Omega)$ , dashed lines—to configuration $\vec{\Omega} = (\Omega,0,0)$ . Thick and thin lines correspond to $\beta = 0.05$ , $\beta = 0.5$ (a) and $\beta = 0.9$ , $\beta = 0.99$ (b).

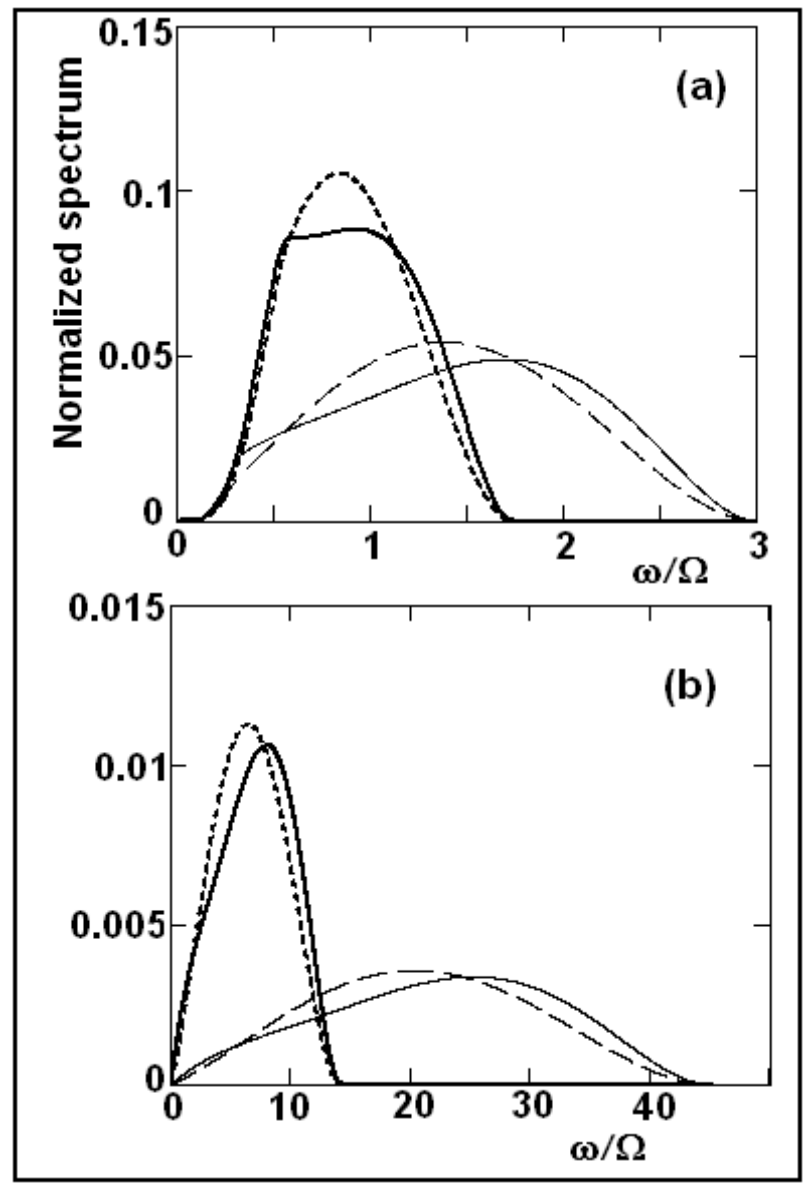

Fig. 3 Normalized frequency spectrum of nonthermal radiation, depending on $\beta$. Solid lines correspond to configuration $\vec{\Omega} = (0,0,\Omega)$, dashed lines—to configuration $\vec{\Omega} = (\Omega,0,0)$. Thick and thin lines correspond to $\beta = 0.5$, $\beta = 0.8$ (a) and $\beta = 0.99$, $\beta = 0.999$ (b).

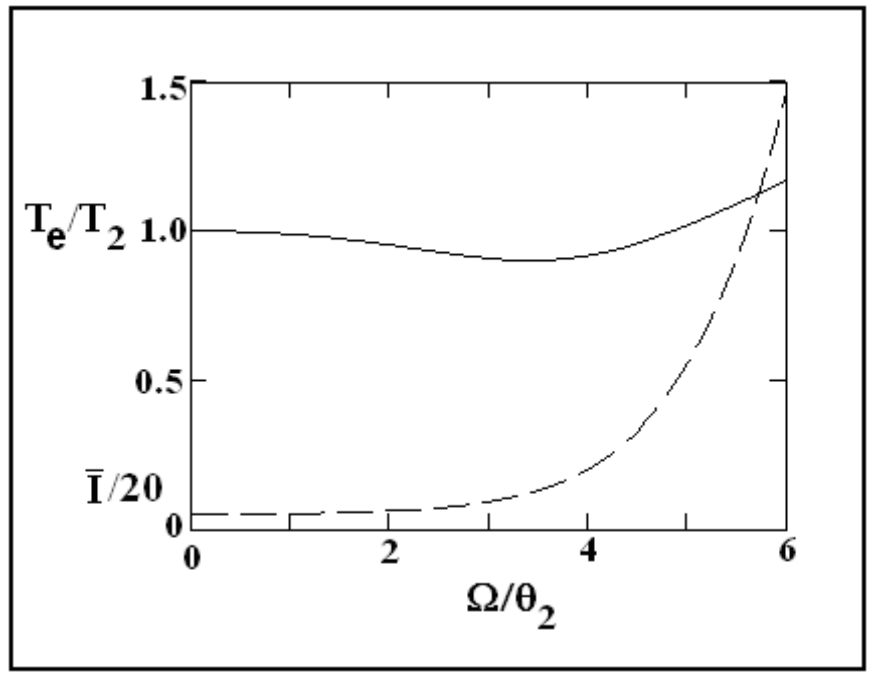

Fig. 4. Effective equilibrium temperature of a motionless particle, $V = 0$ (solid line) and the intensity of radiation normalized onto the factor $I_0$ (see the text) corresponding to configuration in Fig. 1b.

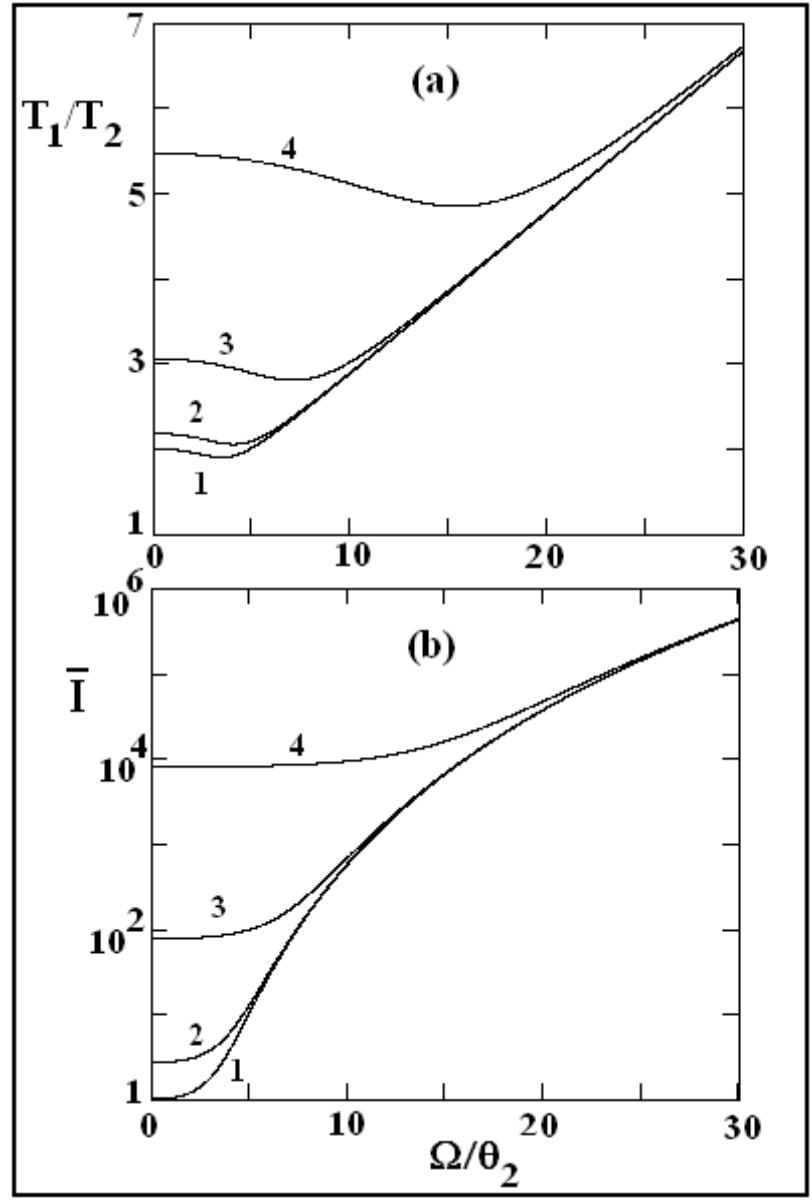

Fig. 5. Same as in Fig. 4 at different velocity factor of the particle. Lines 1 to 4 correspond to $\beta = 0, 0.5, 0.9, 0.99$